\documentclass[twocolumn,tighten,times]{aastex62}
\usepackage{graphicx,xfrac,amsmath,longtable,color}
\usepackage{float}
\usepackage[caption = false]{subfig}
\usepackage{rotating}
\def\la{\mathrel{\hbox{\rlap{\hbox{\lower4pt\hbox{$\sim$}}}\hbox{$<$}}}}
\def\ga{\mathrel{\hbox{\rlap{\hbox{\lower4pt\hbox{$\sim$}}}\hbox{$>$}}}}

\def\kms{km~s$^{-1}$}

\def\dm15{{$\Delta$}$m_{15}$}

\def\v10{$V_{10}$(Si~II)}

\def\W575{$W(5750)$}
\def\W610{$W(6100)$}
\def\6100{the 6100~\AA\ absorption}

\def\msun{M$_\odot$}

\def\gcm3{g~cm$^{-3}$}
\def\cm2g{cm$^{2}$~g$^{-1}$}

\def\CaII7291{[Ca~{\sc II}] $\lambda\lambda$7291,7323}

\def\OI6300{[O~{\sc I}] $\lambda\lambda$6300,6364}

\def\apj{ApJ}
\def\apjl{ApJL}
\def\apjs{ApJS}
\def\aj{AJ}

\def\mnras{MNRAS}
\def\aap{A\&A}

\def\nar{New Astron Rev}

\usepackage[T1]{fontenc}
\usepackage{ae,aecompl}
\newcommand{\blue}[1] {\textcolor{black}{#1}}
\newcommand{\red}[1] {\textcolor{black}{#1}}
\newcommand{\green}[1] {\textcolor{black}{#1}}

\received{2020 July 1}
\submitjournal{ApJ}

\shorttitle{Betelgeuse Merger} 
\shortauthors{Sullivan et al.}

\begin{document}

\title{The Betelgeuse Project III: Merger Characteristics}

\author{J. M. Sullivan}
\affiliation{Department of Astronomy, University of California, Berkeley, California}

\author{S. Nance}
\affiliation{Department of Astronomy, University of California, Berkeley, California}

\author{J.\ Craig Wheeler}
\affiliation{Department of Astronomy, University of Texas at Austin, Austin, Texas}

\correspondingauthor{James M. Sullivan}
\email{jsull3@utexas.edu}

\begin{abstract}

We previously proposed that Betelgeuse might have been spun up by accreting a companion of about 1 \msun. Here we explore in more detail the possible systematics of such a merger and a larger range of accreted masses. We use the stellar evolutionary code \textsc{mesa} to add angular momentum to a primary star in core helium burning, core carbon burning, or shell carbon burning. Our models provide a reasonable ``natural" explanation for why Betelgeuse has a large, but sub-Keplerian equatorial velocity. They eject sufficient mass and angular momentum in rotationally-induced mass loss to reproduce the observed ratio of the equatorial velocity to escape velocity of Betelgeuse, $\approx 0.23$, within a factor of three nearly independent of the primary mass, the secondary mass, and the epoch at which merger occurs. Our models suggest that merger of a primary of somewhat less than 15 \msun\ with secondaries of from 1 to 10 \msun\ during core helium burning or core carbon burning could yield the equatorial rotational velocity of $\sim 15$ \kms attributed to Betelgeuse. For accreting models, a wave of angular momentum is halted at the composition boundary at the edge of the helium core. The inner core is thus not affected by the accretion of the companion in these simulations. Accretion has relatively little effect on the production of magnetic fields in the inner core. Our results do not prove, but do not negate that Betelgeuse might have ingested a companion of several \msun.


\end{abstract}



\section{Introduction}
\label{intro}

Betelguese ($\alpha$ Orionis) is a nearby, massive red supergiant 
(RSG) that provides clues to a broad range of issues of the evolution and explosion of massive stars. It has been difficult to obtain tight constraints on the evolutionary state of Betelgeuse and hence when it might explode and information about the internal rotational state and associated mixing. It is thus important to understand Betelgeuse in greater depth. The recent extreme dimming episode has only added impetus to this quest \citep{Guinan20,Levesque20,Harper20,dharma20}.

The distance to Betelgeuse has been known to only 20\% ($D \approx 197\pm45$ pc; Harper et al. 2008, 2017) a situation that is not improved by {\sl Gaia} that saturates on such a bright star. Key 
properties such as radius and luminosity are thus somewhat uncertain. Within this uncertainty, models of Betelgeuse might be brought into agreement with observations of $L$, $R$, and $T_{eff}$ at either the minimum--luminosity base of the giant branch or at the tip of the red supergiant branch (RSB). By invoking the constraint of the 412 d pulsation \citet{Joyce20} show that this corresponds to the fundamental mode and derive new constraints on the radius and hence the distance and parallax of Betelgeuse, reducing the uncertainty to about 10\% ($D \approx 165^{+16}_{-8}$ pc). This restricts Betelgeuse to be near the tip of the RSG and in core Helium burning. The proximity of Betelgeuese allows other key measurements since its image can be resolved \citep{Habois09}. 

A particularly interesting potential constraint on Betelgeuse obtained with spatially-resolved spectra is the equatorial rotational velocity ($\sim 15$~\kms) measured with $HST$ \citep{Dupree87} and ALMA \citep{Kervella18}. \blue{Our models of Betelgeuse give a critical Keplerian velocity of $\sim 65$~\kms; the observed rotational velocity is thus a substantial fraction of the escape velocity.} In the first paper of {\sl The Betelgeuse Project} \citep{Wheeler17}, we showed that single star models have difficulty accounting for the rapid equatorial rotation and suggested that Betelgeuse might have merged with a companion of about 1 \msun\ to provide the requisite angular momentum to the envelope. In paper II of the series \citep{Nance18}, we explored the possibility of gleaning an understanding of the interior structure of Betelgeuse in particular and RSG in general with some of the techniques of asteroseismology. In this work, we return to the question of whether Betelgeuse might have merged with a companion, as many O and B stars are argued to do \citep{Sana12, deMink14, Dunstall15, Costa15, Renzo19, Zapartas19}. Of primary interest is how and under what circumstances, the merged system ends up rotating at $\sim 23$\% of the critical velocity.

While the hypothesis that Betelgeuse might have merged with a companion is credible and consistent with the {\sl a priori} estimate that Betelgeuse has a probability of $\sim 20$\% of being born in a binary system \citep{deMink14}, it raises a number of interesting issues involving common envelope evolution, the fate of the companion and its angular momentum, and effect on the structure 
of the primary. A main sequence companion of about a solar mass would have a mean density of about 1 \gcm3. That density is characteristic of the base of the hydrogen envelope in the RSG models we consider here, implying that a companion might not be dissolved until it reached the edge of the helium core. If the 
companion merged with the core, the evolution of the primary might be severely altered by anomalous burning and mixing effects, and surface abundances might be affected. The luminosity of an evolved massive star is typically a function of the mass of the helium core and rather independent of the mass of the envelope. If a companion merged with the core of Betelgeuse, then the current luminosity may be a measure of the core mass ($\sim$ 5 to 6~\msun), but the mass of the envelope would be rather unconstrained and probably smaller than the estimates given based on single--star models that attempt to reproduce the luminosity, radius and effective temperature. If there were a coalescence, there would be some mass ejected. \citet{Wheeler17} discussed the possible ramification of such a mass ejection and the possible connection to various structures in the immediate environment of Betelgeuse. 

In \S2 we present the calculations we have done and discuss the 
strengths and weaknesses of attempting to simulate a stellar merger
with a spherical code. Our results are presented in \S3 and a
discussion is given in \S4. 
As this work was nearing completion, \citet{chatz20} presented a somewhat similar analysis and conclusions. Based on their pulsational analysis, \citet{Joyce20} also conclude that Betelgeuse underwent a merger. 

\section{Computations}
\label{comp}

We used the stellar evolution code Modules for Experiments in 
Stellar Astrophysics (\textsc{mesa}; \citealt{Paxton11, Paxton13, Paxton15}). The models were run using \textsc{mesa} version 10398, with the {\it pre-ccsn} test-suite model inlist. To keep the variables to a minimum, we considered test-suite prescriptions in \textsc{mesa}: including Schwarzschild convection and an overshoot parameter of $\alpha = 0.2$. For the rotating models, we again chose \textsc{mesa} test-suite values of mechanisms of angular momentum transport and mixing according to the prescriptions of \citet{Heger03} with the efficiency parameters of the individual viscosity and diffusion coefficients equal to unity. We included magnetic effects as treated by the Spruit/Tayler algorithm \citep{spruit02, tayler73} in some cases, but did not include magnetic effects of the magnetorotational
instability \citep{WKC15}. 
\green{We employed the ``Dutch" and ``de Jager'' mass--loss prescriptions with test-suite wind factors for ``hot'' and ``cool'' winds, respectively and the prescription for rotationally-induced mass loss of \citet{Heger2000}}. We used nuclear reaction network {\sl approx21}. 
The inlist we employed is available upon request from the authors. 

While recent versions of \textsc{mesa} have the capacity to treat binary evolution and common-envelope evolution \citep{Paxton15}, we reserve such studies for the future and here treat the problem in a rudimentary way that nevertheless gives some insights to the relevant physical processes. We have not attempted to treat the companion as a corporeal entity, but allow for its effects by adding the relevant mass and associated angular momentum to the outer envelope of the primary. We refer to our computational process as an ``accretion" throughout this work to distinguish it from the more complex behavior of a true merger, while recognizing its limitations that we discuss below.

We computed a range of models to explore the parameter space for a possible merger of a primary Betelgeuse-like star and a lower-mass companion. We considered a range of primaries with limited variation in companion mass to explore the effect of rotation and primary mass on the merger scenario. In this work, we focus on 
two primary masses, 15 \msun \ and 20 \msun,
but a wider range of secondary masses. The goal is to explore the effect of the epoch of accretion and companion mass on the merger scenario. In subsequent discussion we will refer to these models by the mass of the primary and the secondary, for instance to the 20 + 1 model for the 20 \msun\ primary accreting a 1 \msun\ secondary.


We considered various states of rotation of the initial ZAMS models: non-rotating, ``barely" rotating, and rotating at 200 \kms, a modest fraction of the ZAMS Keplerian speed. The barely rotating models were invoked because our prescription for adding angular momentum failed in truly non-rotating initial models. These barely-rotating models functioned computationally, but had so little angular momentum that they were basically equivalent to non-rotating in terms of structure. A significant rotation near the ZAMS can effect the later core rotation structure, but has very little role to play otherwise after the model evolves up the RSG or has a merger. In this work, we only present the barely-rotating
models. 
These models are initialized with an angular speed set to 0.1\% of breakup. They do not attain substantial rotation until they undergo accretion.

In the \textsc{mesa} models presented here, we have explicitly explored possible circumstances under which a RSG like Betelgeuse might rotate at a substantial fraction of current breakup velocity. We anticipated that the results might be asymptotically independent of the accreted mass since the merger product must eject sufficient mass and angular momentum to be stable. We have thus computed models of \red{15 \msun\ and} 20 \msun\ accreting companions of 1, 2, 3, 5, 7, and 10 \msun.

In the spirit of covering parameter space, we have explored some models in which the accretion occurs during the epoch of early core He burning (hereafter CHeB) as the models cross the Hertzsprung gap. While not precluded, this possibility seems less likely than encountering the secondary during the transition up the RSB. We return to this issue in \S\ref{discussion}. Other sets of models invoked accretion during the epochs of core Carbon burning (CCB) and shell Carbon burning (SCB). We thus evolved the primary star from the pre-main sequence up to a cutoff at one of three epochs: CHeB, CCB, or SCB. As a practical matter, consistent epochs of accretion were enforced for CHeB, CCB, and SCB by specifying a fixed central mass fraction threshold in the \textsc{mesa} inlist before adding mass to the model with the {\it inlist\_accreted\_material\_j} inlist. In practice, accretion was triggered for a central helium mass fraction less than about 0.95 for CHeB, for a central carbon mass fraction of less than 0.1 for CCB and for a central mass fraction of oxygen exceeding 0.7 for SCB. At the chosen epoch, we added mass and angular momentum to simulate a merger and then continued the evolution to near core collapse.

An important aspect of the problem is the deposition of the mass and orbital angular momentum of the secondary. In 3D simulations most of the initial secondary angular momentum is deposited in the outer layers of the primary envelope. That feature is captured in our calculations in at least a rudimentary way. At the chosen accretion epoch, we add to the envelope of the primary an amount of mass corresponding to the chosen secondary and an amount of angular momentum corresponding to the orbital angular momentum of the secondary with an orbit corresponding to the radius of the primary at the epoch when we begin accretion. The addition of the mass and angular momentum is done on a timescale that is long compared to the dynamical time, but short compared to the thermal or evolution times of the envelope as an approximation to the spin-up due to the plunge-in of the companion. The mass and angular momentum are added in the outer few zones by engineering the accretion rate. In \textsc{mesa}, accretion is just a variation on mass loss, invoking a change of sign. Our treatment of accretion is thus the same as the test-suite prescription for mass loss in terms of how it is handled numerically in the code. At the chosen epoch, we used the {\it accreted\_material\_j} test-suite functionality to add the companion mass and its orbital angular momentum to the primary star over a very short fraction of the model lifetime. Subsequently, we evolved the post-accretion model to near core-collapse, using an upper limit of silicon mass fraction of 0.1 as the final stopping point. The addition of mass and angular momentum in our treatment takes place at maximum over about 100 y in the models to avoid numerical convergence issues. This timescale has no physical import, but is longer than the expected plunge-in time. Clearly, we are not capturing that process in any quantitative detail. The excess mass is assimilated as \textsc{mesa} readjusts the density profile on the dynamic timescale of the envelope. The spin up induced by the addition of angular momentum enhances the normal wind mass loss (see \S\ref{mass}). Accretion during CHeB deposits the same mass as for accretion in the RSG phase, but less angular momentum due to the smaller assumed orbit at the presumed onset of merging.  


Our treatment is clearly a simple approximation to the complex
three-dimensional reality of the process of common envelope
evolution (CEE) and merger. \citet{Ivanova16} (see also 
\citealt{MorrisPod07, Taam10, Ivanova13, IvanovaJP, Ivanova_rev16, MacLeod18, chatz20}) describe the basic phases of CEE and the mechanisms for treating it in 3D and 1D. There are three stages to the process, each with associated loss 
of mass and angular momentum: 1) a precursor phase when the stars 
begin to interact and co-rotation is lost; 2) a plunge-in phase with a 
large rate of change of orbital separation and a timescale close 
to dynamical, at the end of which most of the mass of the common 
envelope (CE) is beyond the orbit of the companion; and 3) a 
self-regulated slow inspiral of the companion. There are two basic 
endpoints to CEE: formation of a compact binary system and merger. For mergers, \citet{IvanovaPod03} differentiate three outcomes: a quiet merger, a moderate merger, and an explosive merger. Only the former leaves behind an RSG and hence is pertinent to Betelgeuse.

Mass and angular momentum are lost by dynamical interaction, 
outflow driven by recombination, and shrinking of the orbit. 
The slow inspiral often begins with an envelope that is significantly 
reduced in mass and angular momentum. In some cases, recombination 
outflow can eject nearly all the envelope during the slow inspiral.
The exception to these cases of extreme mass loss is when the primary is substantially more massive than the secondary, the case we consider here for many, but not all models. For small secondary masses, the percentage of mass lost in the precursor phase and the plunge-in phase is of order $q$, the mass ratio of secondary to primary. 

In their treatment of a red giant of modest mass (1.8 \msun),
\citet{Ivanova16} find that companions of mass less than 0.10 \msun, 
corresponding to about 5\% of the primary mass, undergo merger. The
time to merger is about 1000 d, long compared to the dynamical
time of the CE but short compared to the thermal or evolutionary
time of the primary. While these results do not necessarily scale with 
mass, this suggests that for many cases of interest here, a companion of about 1 \msun\ undergoing CEE with a primary of about 20 \msun\ 
is likely to quickly undergo merger while sustaining a substantial 
envelope, as Betelgeuse is observed to have. As we will show below, our models retain the RSG envelope even for much more substantial secondary mass.

The plunge-in phase is expected to induce very asymmetric structures 
and the slow inspiral to yield appreciable departures from spherical symmetry that can be simulated in 3D but are beyond the capacity of 1D models. In 3D there is a significant density inversion in the vicinity of the companion and rather little material near the center of mass of the binary. On the other hand, the 3D simulations often treat the companion star and the red giant core as point sources. In 1D, the primary core, at least, can be modeled in more detail. A 1D code like \textsc{mesa} conserves energy and angular momentum within expected numerical accuracy. \textsc{mesa} also automatically handles energy released by recombination as the envelope expands and the angular momentum lost in winds. In some 1D simulations of CEE, the companion is
treated in a ``thin shell" approximation. In the current work, we have
neglected even that attempt to account explicitly for the presence of
the companion. We thus avoid such complications as the orbit of
the primary core and companion about a center of mass and the
motion of that center of mass during the inspiral.

\citet{Ivanova16} argue that energy conservation during the plunge
phase cannot be properly treated in 1D codes. They recommend that 
the CE should be constructed by assuming adiabatic expansion
of the envelope due to the plunge. While not capturing the full 
richness of the problem, our procedure of adding mass and angular 
momentum to the envelope does result in an adjustment of the 
envelope structure that may be some approximation to a more
accurate treatment.

\section{Results}
\label{results}



\subsection{Evolution in the Hertzsprung-Russel Diagram}
\label{HRD}



\begin{figure}
\center
\includegraphics[width=3 in, angle=0]{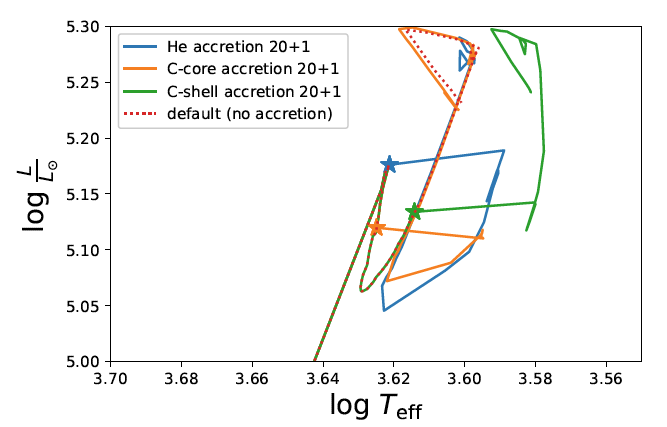}
\caption{Evolution in L and T$_{eff}$ in the late Hertzsprung gap and the RSB for the 20+1 model. The dashed line represents the default model that did not undergo accretion. Stars represent the three epochs of accretion. The blue (uppermost) star and line (nearly indistinguishable from the default model by the end of the evolution) correspond to accretion during core He burning (CHeB) at the base of the RSB. The orange (lowest) star and line (middle track) represent accretion during core C burning (CCB). The green star and line (rightmost star and track) represent accretion during shell C burning (SCB).
\label{HRD20+1}}
\end{figure}

\begin{figure}
\center
\includegraphics[width=3 in, angle=0]{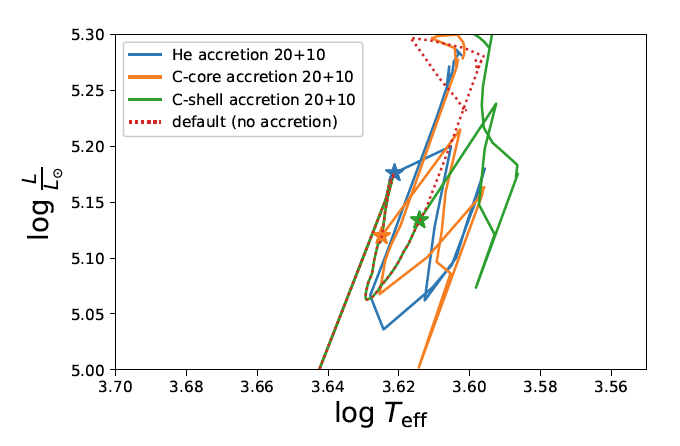}
\caption{Similar to Figure \ref{HRD20+1} but for evolution in the late Hertzsprung gap and the RSB of the 20+10 model. 
\label{HRD20+10}}
\end{figure}

The HRD of all the models are qualitatively similar. Figure \ref{HRD20+1} shows the evolution in the late Hertzsprung gap and RSB for the default model with no accretion and the perturbations on the RSB as mass is added in the three accretion epochs, CHeB, CCB, and SCB, of the 20+1 model. 
The accretion events result in irregular transient loci before settling down to a rather normal evolution up the RSB to the point of collapse of the models. After the transient phase, the models that accreted during CHeB and CCB are nearly indistinguishable from the default model. The model with accretion during SCB ends up in a similar region of the HRD after its post-accretion transient. The CHeB and SCB are, respectively,
slightly brighter and slightly cooler than the default and CCB models at the point of collapse. Figure \ref{HRD20+10} shows similar evolution for the 20+10 model. The larger accreted mass causes more substantial transient perturbations, but modest
difference in the final location of the models at the point of collapse from one another or from the 20+1 model. The 20+10 CHeB model ends up essentially in the same position as the 20+1 model. The 20+10 CCB model has a more complicated path, but ends up only very slightly brighter and hotter than the 20+1 model. It ends slightly brighter and cooler than the 20+10 CHeB model. The 20+10 SCB model ends up 
somewhat dimmer and cooler than the 20+10 CCB model. On the scale plotted, the 20+10 SCB model ends perceptibly hotter than the 20+1 SCB model. 

\subsection{Evolution of Mass}
\label{mass}

\begin{figure}
\center
\includegraphics[width=3 in, angle=0]{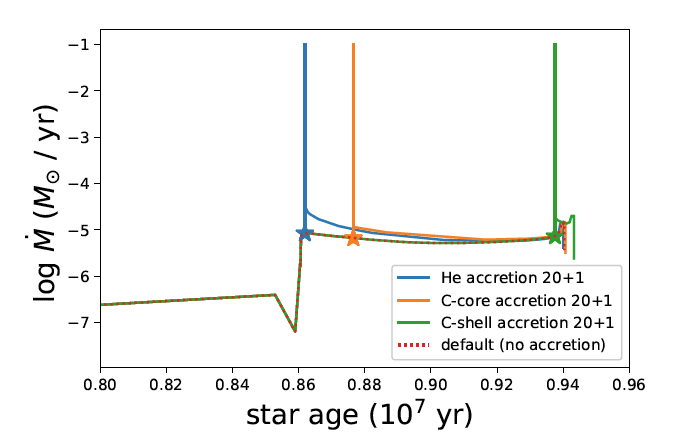}
\caption{The evolution of the mass loss rate from late on the
main sequence to the point of collapse for the 20+1 model. The red dashed line beginning on the left is the default model (this line is identical to and obscured by the line for the SCB model until the very end of the evolution). The blue star and spike (leftmost) correspond to the CHeB model. The orange star and spike (middle features) correspond to the CCB model. The green star and spike (rightmost) correspond to the SCB model. The excursion at $t = 0.86\times10^7$ yr corresponds to the contraction at the Terminal Age Main Sequence.
\label{Mdot20+1}}
\end{figure}

\begin{figure}
\center
\includegraphics[width=3 in, angle=0]{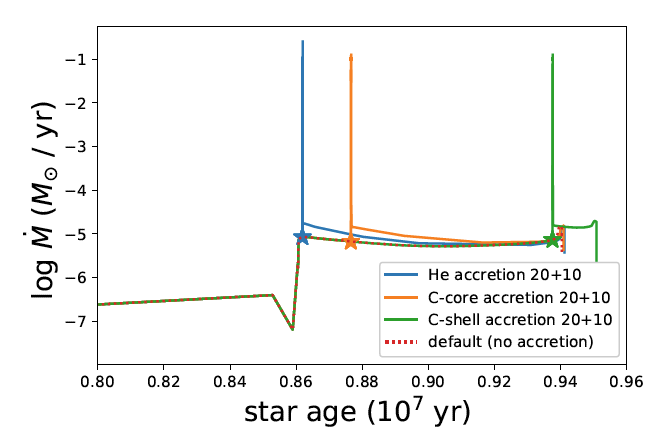}
\caption{Similar to Figure \ref{Mdot20+1}, but for the evolution of the mass loss rate for the 20+10 model.
\label{Mdot20+10}}
\end{figure}

In 3D models, the surface layers are ``shock heated" and quickly ejected prior to the plunge-in. In our models, the associated spin-up of the outer envelope leads to a certain degree of mass loss, that, while perhaps not quantitatively equivalent to a full 3D simulation, captures some of the essence of the interaction \citep{zhaofuller20}. Figure \ref{Mdot20+1} gives the mass-loss history, beginning near the end of the main sequence phase, of the 20+1 model corresponding to the three principal epochs, CHeB, CCB, and SCB. Figure \ref{Mdot20+10} gives similar information for the 20+10 model. At the epoch of accretion, the mass loss rate jumps by about 4 orders of magnitude, abetted by the attempt of the models to restore hydrostatic and thermal equilibrium and by the rotationally-induced mass loss. 
Despite the significant perturbation to the structure of the models, the mass loss rates by the epoch of collapse are very similar for all three accretion epochs and for both the 20+1 and 20+10 models.

\begin{figure}
\center
\includegraphics[width=3 in, angle=0]{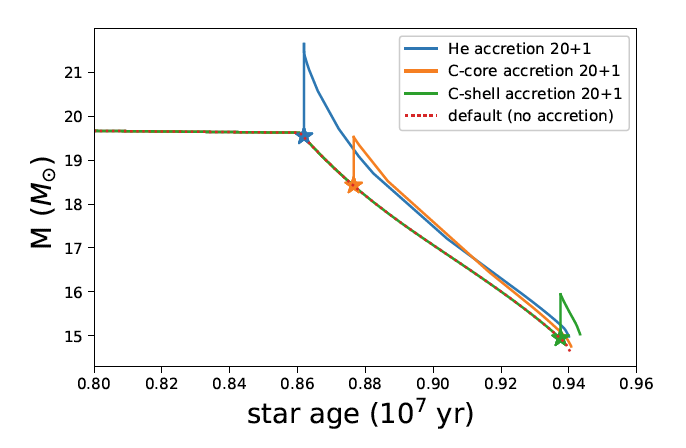}
\caption{The evolution of the mass from late on the main sequence to the point of collapse for the 20+1 model. The red line beginning on the left is the default model. The blue star and spike (leftmost) correspond to the CHeB model. The orange star and spike (middle features) correspond to the CCB model. The green star and spike (rightmost) correspond to the SCB model.
\label{RSGM+1}}
\end{figure}

\begin{figure}
\center
\includegraphics[width=3 in, angle=0]{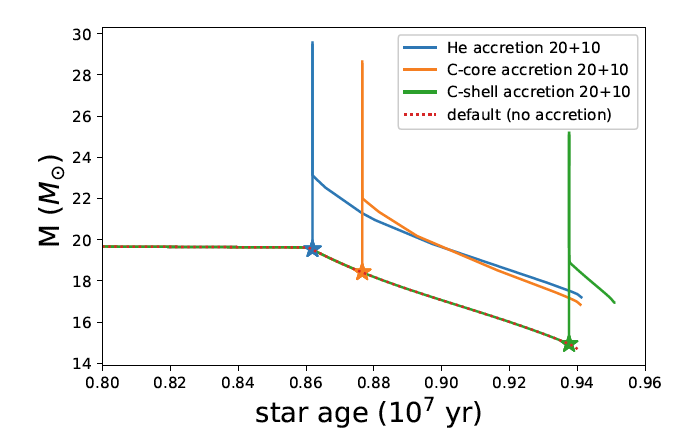}
\caption{Similar to Figure \ref{RSGM+1}, but for the evolution of the mass for the 20+10 model. A color version is available online.
\label{RSGM+10}}
\end{figure}

Figure \ref{RSGM+1} and Figure \ref{RSGM+10} give the mass history for the default, 20+1 and 20+10 models. The excess mass accreted is rapidly expelled in a transient phase of rapid loss of mass and angular momentum as qualitatively expected for the plunge-in phase. The 20+1 model rapidly returns to the same mass and subsequent mass loss rate as the default model for the CHeB and CCB models and the mass of these models is virtually identical at the point of core collapse. The 20+1 SCB model does not have time to relax to the original track before core collapse, but nevertheless ends up with a very similar final mass, $\sim 15$ \msun, as the default, CHeB, and CCB models. The 20+10 models undergo much stronger perturbations. The final masses remain somewhat larger than the 20+1 models for all three epochs of accretion, by about 2 \msun, but nevertheless end up very similar to one another with a mass at core collapse of $\sim 17 - 18$ \msun, nearly the mass of the original ZAMS model. 
\green{Even though the SCB models undergo accretion almost a million years after the CHeB models, they reach nearly the same mass due to more rapid mass shedding between accretion and core collapse for the SCB models.}

Table \ref{massloss} gives the final mass and the total mass ejected from the system during the accretion for ``mergers" with a primary of 20 \msun\ occurring at the three principle epochs, CHeB, CCB, and SCB, and for a range of accreted masses. Note that the final masses, ranging from 15 to 17 \msun, are remarkably independent of the mass of the secondary and the epoch at which accretion occurs. 

The mass lost from the system, ranging from roughly \red{6} to \red{13} \msun\ in Table \ref{massloss} is a combination of the loss of mass accreted plus loss of mass from the primary itself. The latter is due to winds prior to the accretion event and then the rotationally-induced mass loss after the accretion. While it is difficult to isolate the loss of accreted mass from the mass lost directly from the primary, Table \ref{massloss} shows that the net mass loss exceeds the accreted mass so that some mass must be lost from the primary. With the assumption that none of the accreted mass is retained by the primary, the mass lost from the primary is 20 \msun\ minus $M_f$, or 3 - 5 \msun.

\begin{table}
\caption{\green{Final Mass ($M_f$) and Total Mass Ejected ($M_{ej}$) from the System} in \msun\ for Models with ZAMS Mass 20 \msun\ as a function of accreted mass ($M_2$) in \msun. }
\label{massloss}
\begin{tabular}{lccccccc}
\hline
& He core & & C core & & C shell \\
Secondary  & $M_f$ & $M_{ej}$ & $M_f$ & $M_{ej}$ & $M_f$ & $M_{ej}$  \\

\hline
1 & 15.09 & 5.91 & 14.76 & 6.24 & 15.04 & 5.96 \\
\\
2 & 14.98 & 7.02 & 14.56 & 7.44 & 14.37 & 7.63 \\
\\
3 & 14.70 & 8.30 & 14.41 & 8.59 & 14.25 & 8.75 \\
\\
5 & 14.98 & 10.02 & 14.67 & 10.33 & 14.53 & 10.47 \\
\\
7 & 15.66 & 11.34 & 15.36 & 11.64 & 15.45 & 11.55 \\
\\
10 & 17.20 & 12.80 & 16.85 & 13.15 & 16.95 & 13.05 \\

\hline           
              
\end{tabular}
\end{table}



\subsection{Evolution of Angular Momentum and Angular Velocity}
\label{angmom}

During the accretion and redistribution process, some angular momentum 
is lost to the surroundings in the rotation-enhanced wind, and some is 
retained to diffuse inward toward the primary core. 
The angular momentum that is retained is redistributed by an inward diffusive wave of angular momentum. The profiles of the specific angular momentum and angular velocity quickly evolve to stable forms delineated by an inward propagating front with the specific angular momentum increasing outward beyond the front and the angular velocity being nearly constant. 

\begin{figure}
\center
\includegraphics[width=3 in, angle=0]{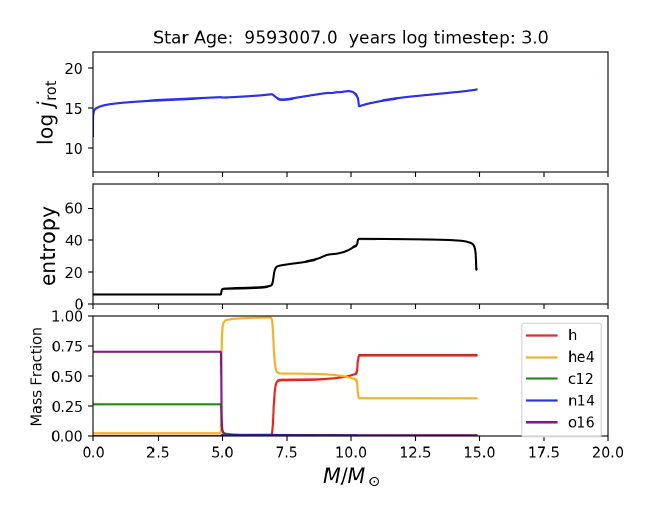}
\caption{The distribution of $j_{rot}$, $s$, and composition just prior to accretion for the 20 \msun\ model in the core carbon burning (CCB) phase.
\label{20jsomegacomppreCCB}}
\end{figure}

\begin{figure}
\center
\includegraphics[width=3 in, angle=0]{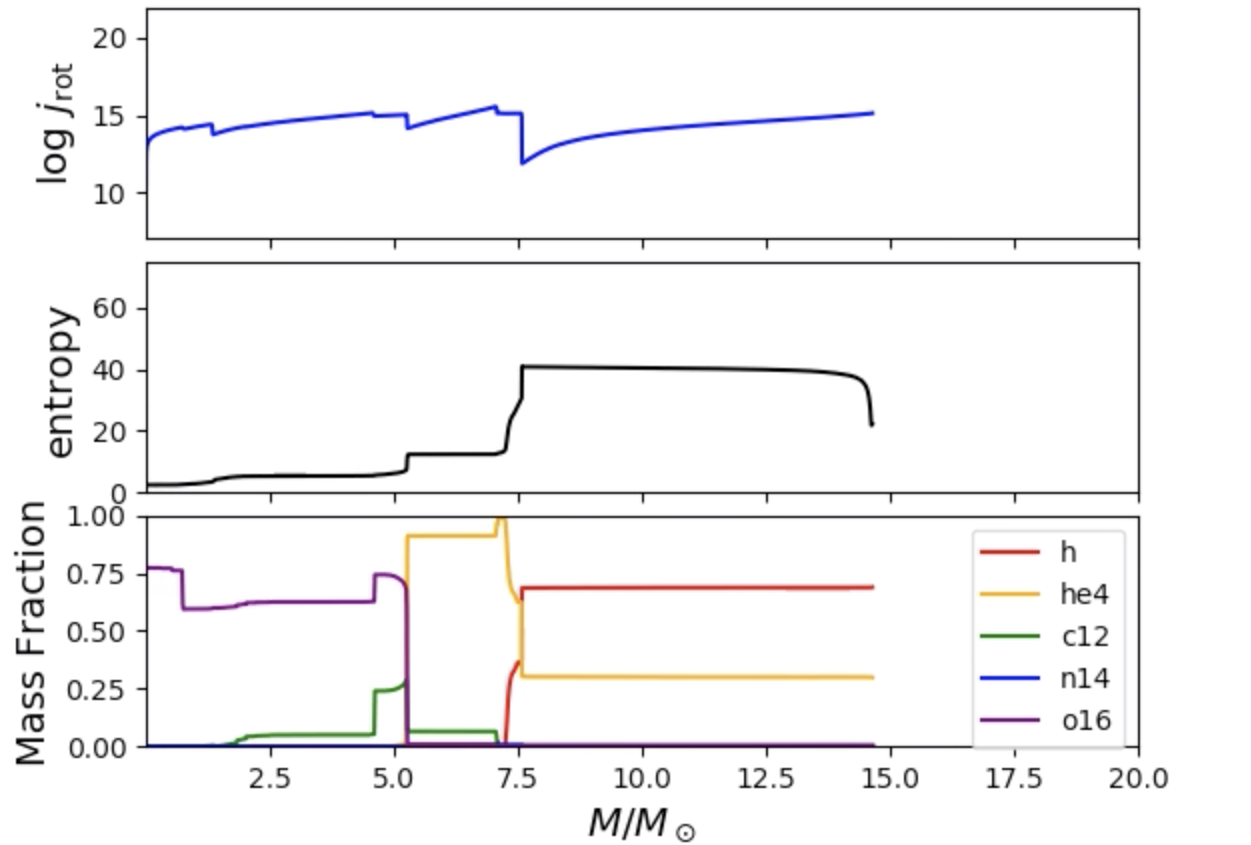}
\caption{The distribution of $j_{rot}$, $s$, and composition
near the point of core collapse for the default 20 \msun\ model that did not undergo accretion.
\label{20jsomegacompdefault}}
\end{figure}

\begin{figure}
\center
\includegraphics[width=3 in, angle=0]{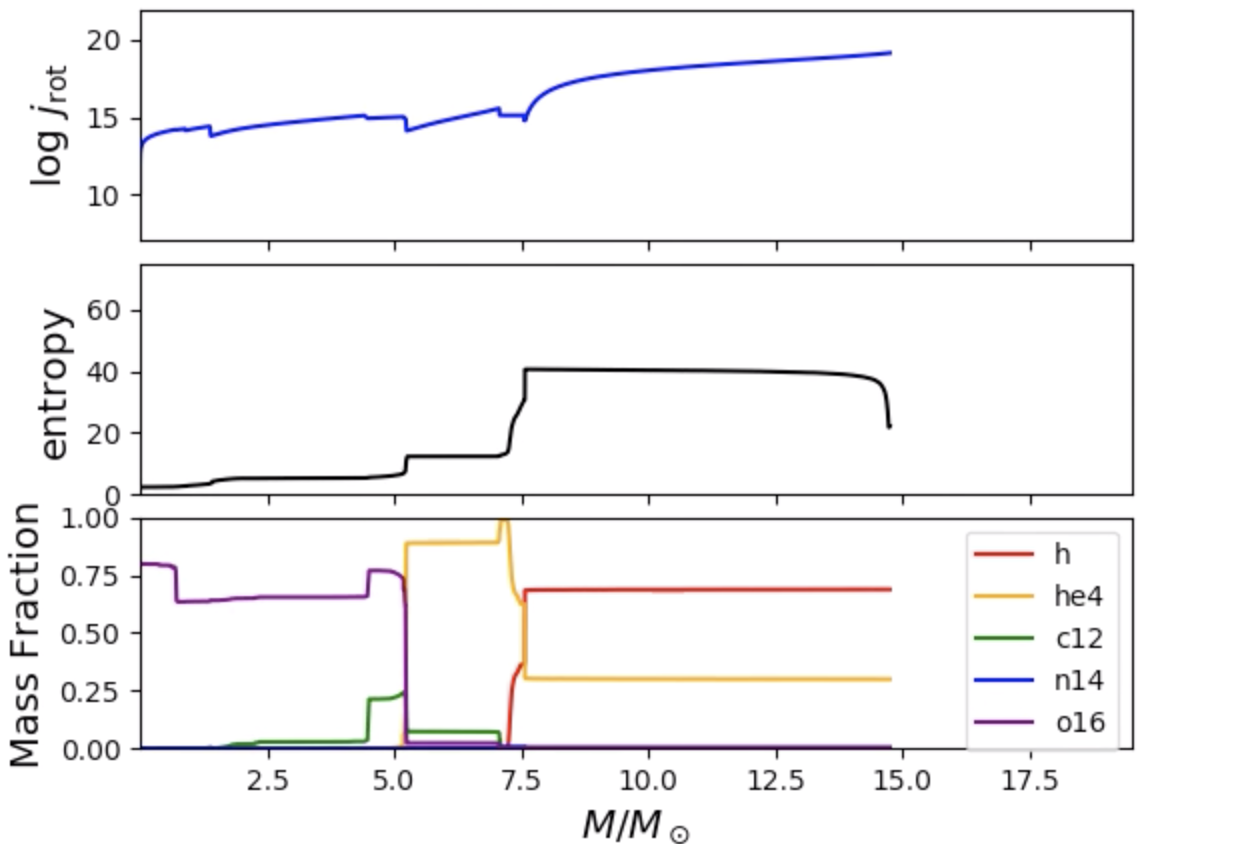}
\caption{The distribution of $j_{rot}$, $s$, and composition
near the point of core collapse for the 20 \msun\ model that accreted 1  \msun\ during core carbon burning (CCB).
\label{20+1jsomegacompCCB}}
\end{figure}

\begin{figure}
\center
\includegraphics[width=3 in, angle=0]{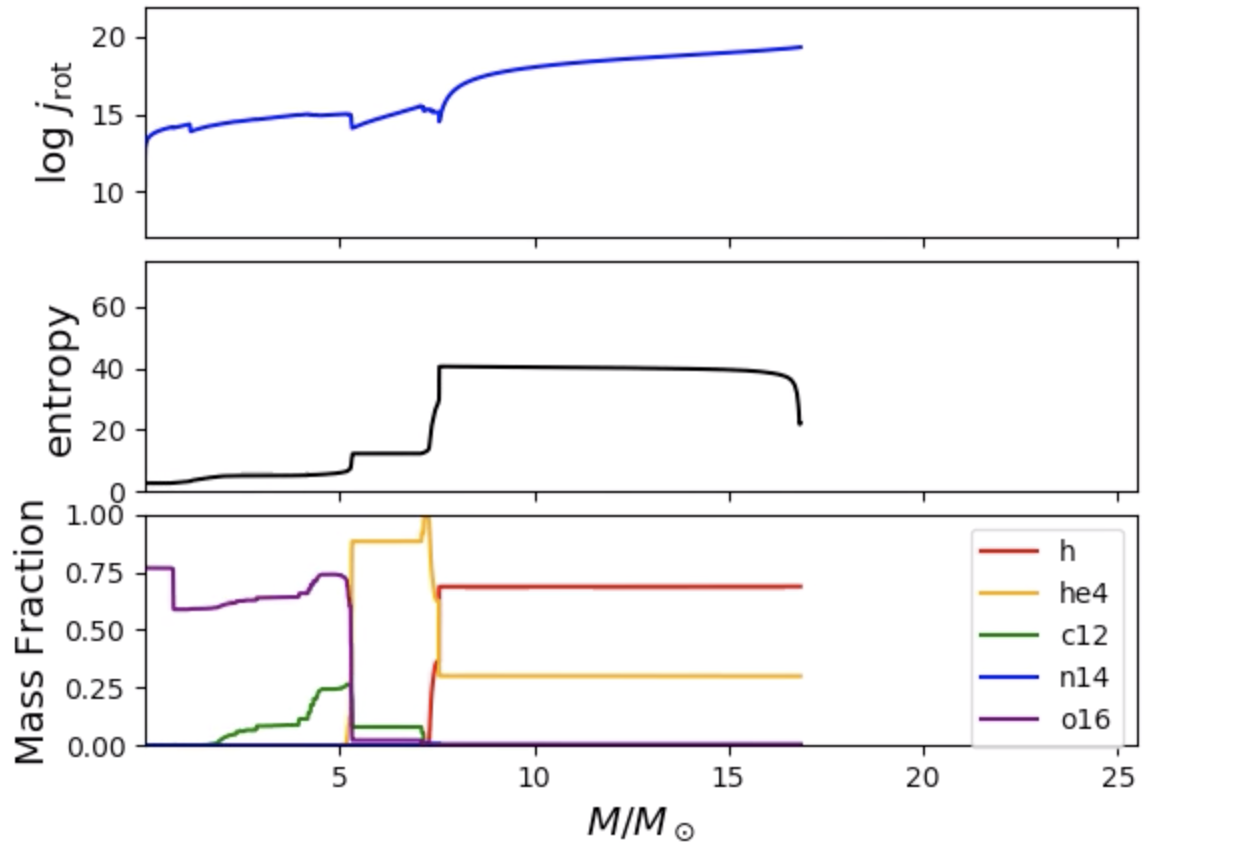}
\caption{The distribution of $j_{rot}$, $s$, and composition
near the point of core collapse for the 20 \msun\ model that accreted \red{10}  \msun\ during core carbon burning (CCB).
\label{20+10jsomegacompCCB}}
\end{figure}

Figure \ref{20jsomegacomppreCCB} gives the distribution with mass of the specific angular momentum, $j$, the specific entropy, $s$, and the composition distribution for the 20 \msun\ model just before accretion in the carbon core-burning phase. Figures \ref{20jsomegacompdefault}, \ref{20+1jsomegacompCCB} and \ref{20+10jsomegacompCCB} give the corresponding distributions near the epoch of collapse for the default model, the 20+1 model and the 20+10 model, respectively. Prior to the accretion phase, the model has an inner homogeneous 5 \msun\ core composed primarily of oxygen, a helium shell extending from 5 to about 7 \msun, a shell composed of roughly 50\% H and He, and the outer RSG envelope. After accretion, the angular momentum per unit mass and the angular velocity in the outer envelope jump substantially. 
A few years after accretion (arbitrarily set by our numerics) the composition distribution is virtually unchanged (there are some changes in detail), but the ingoing wave of angular momentum has propagated to the boundary between the outer envelope and the H/He shell. By the epoch of collapse, the angular momentum distribution in the outer envelope has scarcely changed. The wave of angular momentum has swept through the H/He shell, but is halted at the outer boundary of the He shell at 7 \msun\ for both the 20+1 and the 20+10 models.

All the model have inner regions of negative gradient in $j$ in regions of sharp composition gradients. These must be stabilized against the Rayleigh instability by the associated composition gradients. We have not investigated this condition in detail. 


The net effect is that both the total mass of the model and the mass of the inner core are scarcely changed whether 1 or 10 \msun\ is accreted. All the models, CHeB, CCB, and CSB, end up with an oxygen core of about 5 \msun\ and the mass of helium and heavier elements beneath the envelope of about 7 \msun. The final angular momentum distribution of the outer envelope is very similar for the 20+1, and the 20+10 models.  \green{Both accreting models have a substantially larger envelope angular momentum than the default model.}

Figures \ref{20+1jsomegacompCCB} and \ref{20+10jsomegacompCCB} show, however, that the inner core composition structure \green{near core collapse} is somewhat different.  Inspection of the models shows that there is little difference among the final models for the default model and the 20+1 and 20+10 CHeB models. \green{The structure of the helium-rich shell is virtually identical. There are small quantitative differences in distribution of the oxygen in the inner oxygen core. There are also small differences in the extent of the carbon-rich outer layer of the oxygen core that lies immediately beneath the helium shell, with slightly thicker shells for the accreting models. The CCB and CSB models show similar behavior. Most of the models show a small carbon abundance but virtually no oxygen in the helium shell. The exceptions are the CSB models. The 20+1 CSB model shows a small oxygen abundance equivalent to that of carbon in the helium shell. The 20+10 CSB model reveals an enhanced abundance of both carbon and oxygen in the helium shell compared to the default model, with $ \green{36}
$\% carbon and $ \green{28}
$\% oxygen by mass at the expense of helium which has decreased to $ \green{49}
$\%. }
The implication is that the inner composition structure of Betelgeuse could be rather different depending on the mass accreted with basically no indication reflected in the outer, directly observable structure. 

\citet{Ivanova16} presented a model of a primary of 1.8 \msun\ and secondary of 0.1 \msun (model M10; their figure 7). While the mass scale is smaller than we consider here, the mass ratio for the 20+1 models, $\sim 0.05$, is about the same. 
The angular velocity as a function of mass 50 days after the plunge-in is basically flat throughout the model. The value of the angular velocity, $\sim 3\times 10^{-7}$ rad s$^{-1}$ is interestingly, if fortuitously, close to the value we find. The peak value of the angular momentum per unit mass is about a factor of 30 less than we find. The flat angular velocity profile in the 3D simulations 
seems to arise naturally in our MESA simulations. 

The significant departures in behavior between model M10 and
the results we present in Figures \ref{20+1jsomegacompCCB} and \ref{20+10jsomegacompCCB} are found in the innermost and the outermost regions. \citet{Ivanova16} do not consider the inner core, so they do not explore the distribution of angular momentum we depict in the core. On the other hand, \citet{Ivanova16} find a distinct decrease in both the specific angular momentum and the angular velocity in the outer 0.1 \msun\ of their models that our models do not reveal. This difference probably arises in the loss of mass and angular momentum in the dynamical plunge-in phase that we do not treat accurately. 
While we are clearly not reproducing the interaction and plunge-in and associated angular momentum ejection properly, we do seem to capture many of the major qualitative aspects of the acquisition and redistribution of angular momentum due to merger.

\subsection{Evolution of Entropy}
\label{entropy}

\citet{Ivanova16} give an extensive discussion of the treatment of entropy in CEE. 
They argue that 1D stellar codes should add the energy as mechanical energy rather than ``heat" that moves the material to a higher adiabat. 
The entropy determines the location at which the recombination energy is able to overcome the binding energy. For this reason, the entropy generation computed in 1D codes is likely to predict different outcomes for 1D rather than 3D CE evolution. \citet{chatz20} find relatively little heating effects in their 3D merger simulation of Betelgeuse. We note, however, that heating during merger can lead to non-linear envelope pulsations and to potentially large mass loss \citep{Clayton17}. This aspect is beyond the scope of the current paper, but deserves closer attention.

Since we do not explicitly treat the secondary, we cannot address many of these issues directly, but we can examine the behavior of the entropy in our models to see where our models agree or disagree with other treatments. We neglect the generation of entropy in the merger and plunge-in phase, but our simulations can in principle produce some shear dissipation and entropy in the outer layers by treating the effective diffusion constant and viscosity associated with the Kelvin-Helmholtz instability. We may also capture the generation of some entropy from the flattening of the rotational profile. 

Inspection of our models (see Figures \ref{20jsomegacompdefault}, \ref{20+1jsomegacompCCB} and \ref{20+10jsomegacompCCB}) show that the way we have treated the problem, there is very little perturbation to the entropy of the outer layers. The specific entropy has almost the same value before and after the accretion phase and until the epoch of core collapse. 

\subsection{Recombination}
\label{recombination}

Hydrogen and helium recombination 
can help to trigger envelope instability depending on where and when the energy is released. The time-scale of recombination runaway can be up to several hundred days and gets longer as the mass of the companion decreases. In such cases, radiative losses can become important so that 3D simulations that lack that feature are no longer appropriate. For all their limitations, 1D codes like MESA can handle this aspect of the physics. 

In our calculations, we just add angular momentum, not heat. This results in a change in the distribution of kinetic energy in the envelope that is redistributed as the calculation proceeds. If this process leads to expansion of part of the envelope, triggering recombination, then \textsc{mesa} should compute the recombination self-consistently. It is not clear that this method properly captures the reality that would accompany the full 3D situation with radiative losses and recombination. We consider cases where the accreted mass is modest that might correspond to long recombination timescales, but also accretion of considerable mass. For stability, our numerical process requires that the mass be added on timescales that may be long compared to expected recombination timescales. Given our somewhat artificial means of adding mass and angular momentum, and the transient large perturbation to the envelope structure when large masses are accreted, \textsc{mesa} should compute the recombination self-consistently in the transient adjustment phase as hydrostatic equilibrium is maintained and thermal equilibrium is restored.


\subsection{Magnetic Fields}
\label{mag}

As noted in \S\ref{comp}, we included magnetic effects as treated by the \textsc{mesa} Spruit/Tayler algorithm in some cases, but did not include magnetic effects of the magnetorotational instability \citep{WKC15}. The omission of the latter will undoubtedly alter the quantitative, if not qualitative results. The Spruit/Tayler mechanism gives results that typically weight the radial component, $B_r$, orders of magnitude less than the toroidal component, $B_\phi$. The magnetorotational instability tends to give the radial component about 20 per cent of the toroidal component. Another important caveat is that \textsc{mesa} computes the magnetic field structure based on the instantaneous structure of the model. In reality, the field only decays on a dissipation timescale that might in some circumstances be long compared to the evolutionary timescales. This would lead to fossil magnetic field in a region that made a transition from being unstable to stable to the Spruit/Tayler instability. \textsc{mesa} has no means to treat the existence and decay of such fossil fields. The magnetic structure we compute is thus interesting, but should not be given any quantitative weight. 

\begin{figure}
\center
\includegraphics[width=3 in, angle=0]{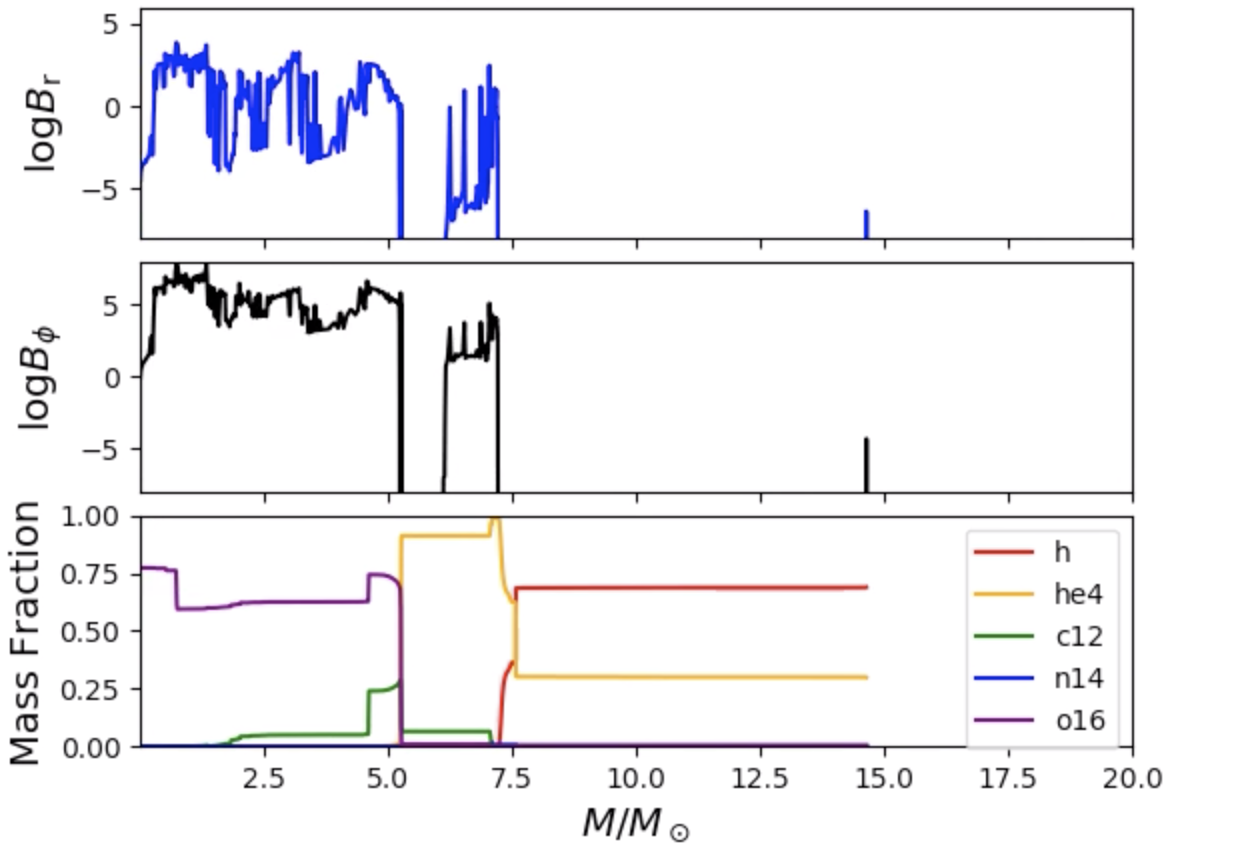}
\caption{The distribution of radial (top panel) and toroidal (middle panel) magnetic field, and composition (bottom panel) near the point of core collapse for the model with 20 \msun\ that did not undergo accretion.
\label{Bdefault}}
\end{figure}

\begin{figure}
\center
\includegraphics[width=3 in, angle=0]{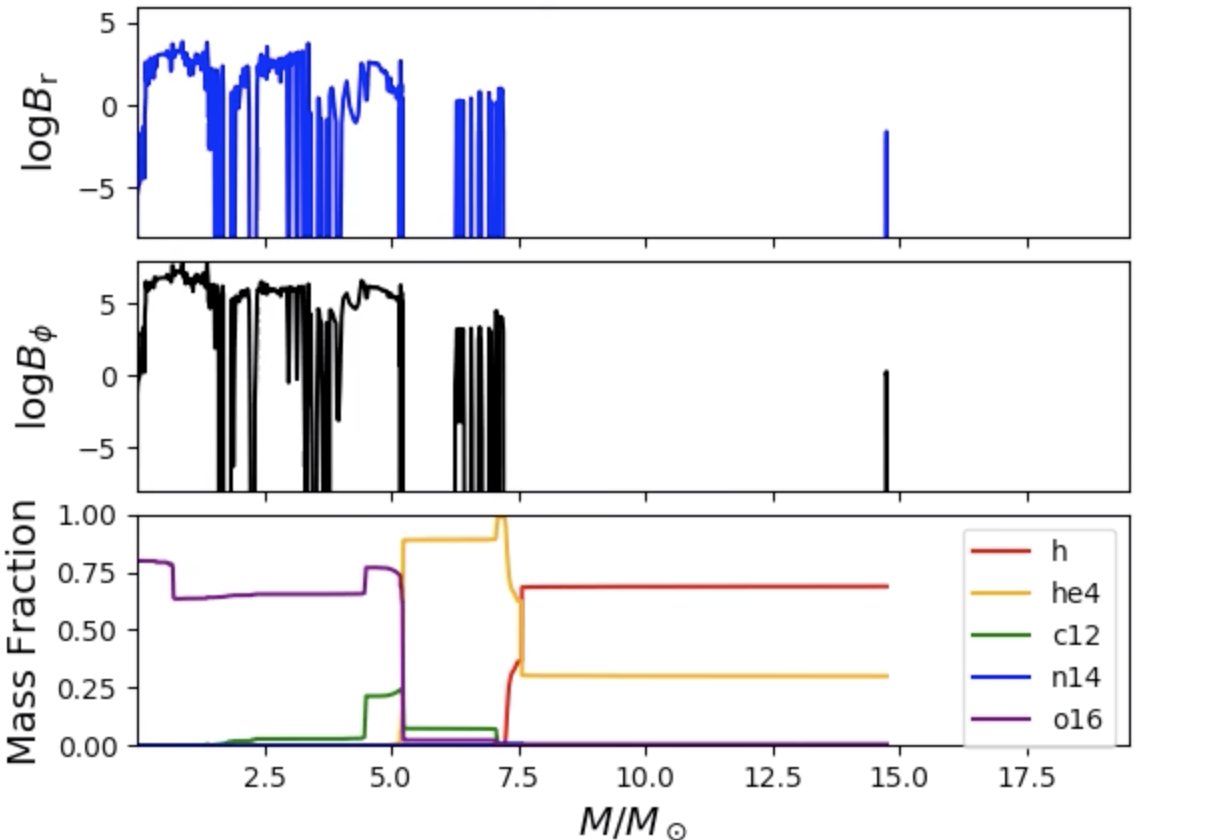}
\caption{\green{Similar to Figure \ref{Bdefault} but} for the model with 20 \msun\ accreting 1 \msun\ of mass during central carbon burning.
\label{B1}}
\end{figure}

\begin{figure}
\center
\includegraphics[width=3 in, angle=0]{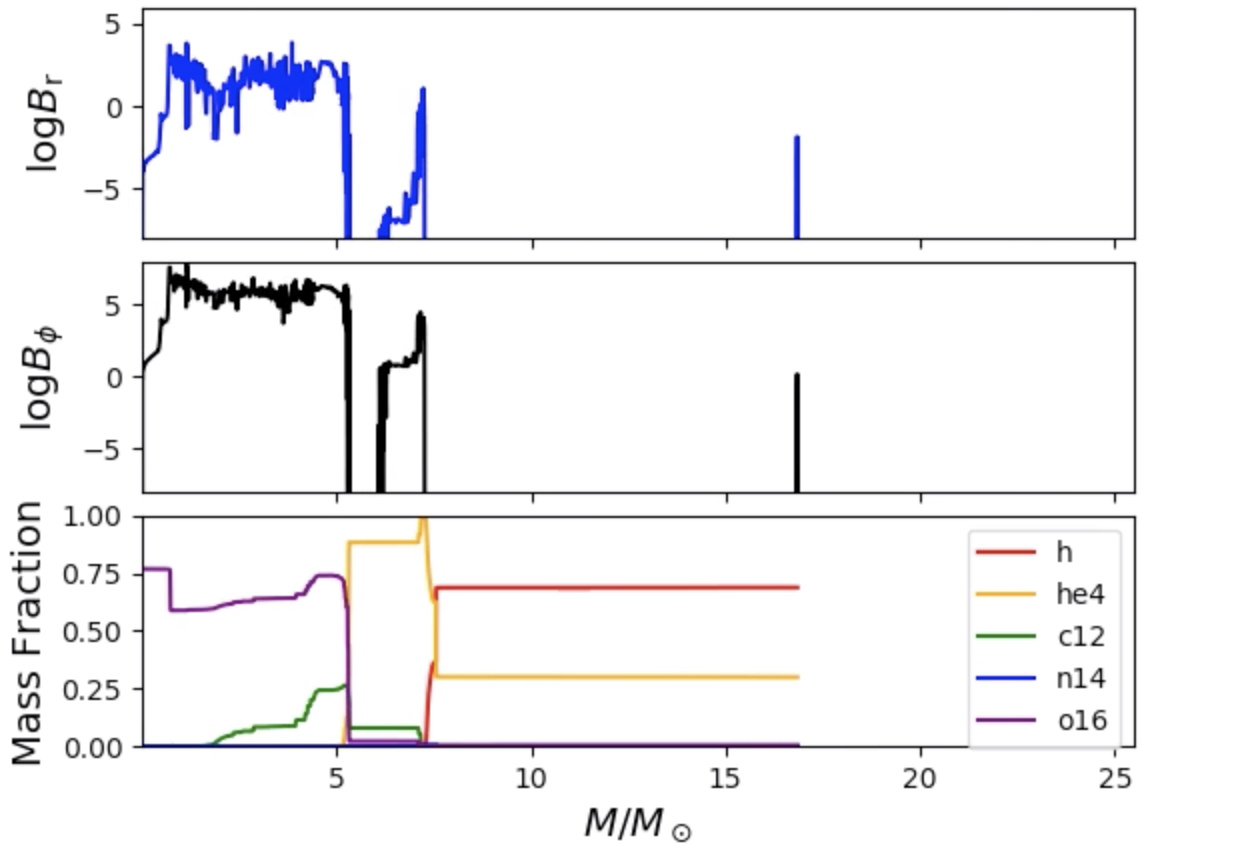}
\caption{\green{Similar to Figure \ref{Bdefault} but} for the model with 20 \msun\ accreting 10 \msun\ of mass during central carbon burning.
\label{B10}}
\end{figure}

\green{Figures \ref{Bdefault}, \ref{B1}, and \ref{B10} show the final field configuration for the default model and the 20+1 and 20+10 models that accreted during core carbon burning, respectively. }
By the end of the calculation, the accreting models showed spikes of modest field, $\sim 1$ G in both $B_r$ and $B_\phi$, in the very outer layers where the models show a generic negative gradient in specific entropy. \green{The default model showed similar spikes, but of considerably smaller amplitude.} Of more interest is the substantial and complex field distribution in the inner core generated as various components burn out and the core contracts and spins up generating shear and magnetic fields. 
\green{These effects would be} amplified if the initial rotation were larger on the ZAMS than we assume here. 
\green{All three models show a more substantial field in the outer part of the helium shell, reaching up to the base of the hydrogen envelope. The peak fields are of order 1 G and 1000 G for the radial and toroidal fields, respectively, with considerable variation with radius that is likely to be affected by issues of numerical resolution. All three models then have an inward gap where the fields are very small. The fields are then large, but variable, in the innermost layers of the oxygen core.  The radial fields peak at $\sim$ 1000 G and the toroidal fields at $\sim 10^6$ to $10^7$ G. In these models the fields peak off center and the toroidal field declines to about 1 G in the center. The accretion appears to have a quantitative, but not qualitative effect on the field strength and distribution just prior to collapse. Subsequent core collapse by a factor of $\sim 100$ in radius would amplify the field by compression alone by a factor of $\sim 10^4$. The resulting field of $\sim 10^{11}$ G would not be dynamically significant, but would give ample seed field for growth of the field in the proto-neutron star by the MRI \citep{Akiyama03, Obergaulinger09, Moesta18}  }

\subsection{\blue{Insensitivity of Final Equatorial Velocity to Accreted Mass }}
\label{finalv}

The original motivation of \citep{Wheeler17} for hypothesizing that Betelgeuse might have merged with a companion was the difficulty of accounting for the nominal currently-observed equatorial rotation velocity, $\sim15$ \kms, allowing for inclination. A companion mass of $\sim 1$ \msun\ was estimated from simple arguments based on conservation of angular momentum. 

If a merger occurred in Betelgeuse, the product must have settled into a state for which the rotation is sub-Keplerian. This global criterion is independent of the masses of the primary and secondary involved in the merger. The implication is that the loss of mass and angular momentum must adjust to meet this criterion rather independently of the masses involved and the epoch of accretion. To explore this notion, we investigated the mergers of a range of primary and secondary masses. Here we concentrate on primaries of 15 and 20 \msun, but consider a range of secondaries up to a rather extreme 10 \msun. 

\begin{figure}
\center
\includegraphics[width=3.5 in, angle=0]{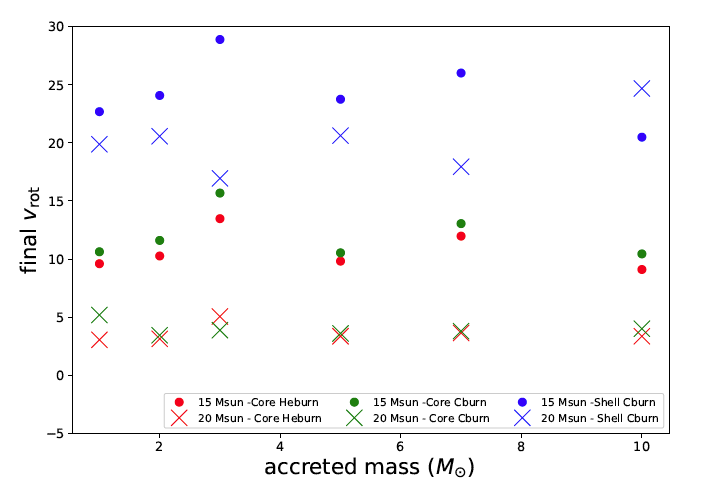}
\caption{The equatorial rotational velocity \green{as a function of} the accreted companion mass for the set of models containing 15 and 20 \msun\ primaries at Helium burning (red), core Carbon burning (green), and shell Carbon burning (blue).}
\label{vrotbare}
\end{figure}

\begin{figure}
\center
\includegraphics[width=3.5 in, angle=0]{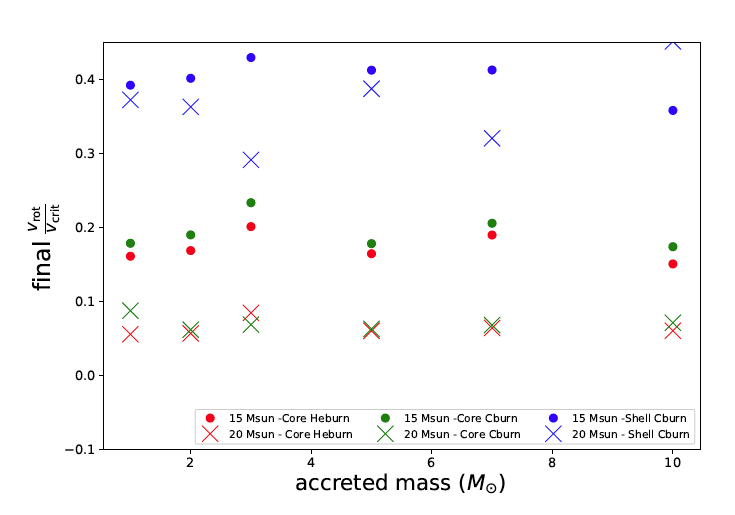}
\caption{The ratio of equatorial rotational velocity to the critical equatorial rotational velocity \green{as a function of} the accreted companion mass for the set of models containing 15 and 20 \msun\ primaries at Helium burning (red), core Carbon burning (green), and shell Carbon burning (blue).}
\label{vrotcrit}
\end{figure}

\blue{In Figures \ref{vrotbare} and \ref{vrotcrit}, we explicitly compare the final equatorial rotation velocity and its ratio with the critical rotation velocity, respectively. We find that, broadly, the final rotational velocities of the models were \green{rather} independent of the companion mass accreted. \green{For} a given accretion epoch, the final rotational velocities for the 15 \msun\ primary models were typically higher than those of the 20 \msun\ primary models. \green{The results for the CHeB and CCB models were very similar. The final velocity for the SCB models were substantially higher, presumably due to the smaller time from accretion to collapse that prevented more loss of mass and angular momentum (see Figures \ref{RSGM+1} and \ref{RSGM+10}). } }


\green{Taking the results of our models at face value and interpolating in Figure \ref{vrotbare}, the rotational velocity attributed to Betelgeuse, $\sim 15$ \kms, could be reproduced by a model with a primary of somewhat less than 15 \msun\ accreting between 1 and 10 \msun\ in the CHeB and CCB epochs. This velocity might also be attained by accreting any of a broad range of masses onto a primary of somewhat more than 20 \msun\ in the later SCB epoch. Accretion of as much as 10 \msun\ at the SCB epoch would require an even more massive primary. Similar conclusions are reached by examination of Figure \ref{vrotcrit} where the ``observed" ratio of equatorial rotational velocity to the critical equatorial rotational velocity is $\sim 0.23$.
}

\section{Discussion and Conclusions}
\label{discussion}


We have used \textsc{mesa} to approximately simulate the merger of a massive primary with secondaries of a range of masses in an attempt to better understand the apparent equatorial rotation velocity of Betelgeuse, $\sim 15$ \kms, $\approx 0.23$ of the Keplerian velocity. We simulate a merger by adding to the envelope of the primary the mass of the secondary and angular momentum corresponding to the orbital angular momentum of the secondary at the radius of the primary at the epoch when we begin accretion. We consider accretion during core helium burning, core carbon burning and shell carbon burning and compute the resulting models to near core collapse. \citet{Joyce20} conclude that Betelgeuse merged prior to the later carbon-burning phases. Our core helium burning models might thus be most pertinent to Betelgeuse, but our other models might pertain to other cases of RSG merger. We discuss the limitation of tackling a manifestly 3D problem with a 1D code, including treatment of the entropy distribution in the envelope and the effects of recombination on the energetics of envelope loss. 

While the final mass of the primary and the equatorial velocity depend somewhat on circumstances, we find that with the assumptions we have made they are remarkably insensitive to the mass of the secondary and the epoch of accretion. For a 20 \msun\ primary, the final mass is \green{ 15 - 20} \msun, nearly independent of the mass of the secondary or the epoch of accretion. The equatorial velocity is consistent with the observed value within a factor of a few. The results for accreting 1 \msun\ are not drastically different than those for accreting 10 \msun. \green{ Our models suggest that the rotation of Betelgeuse could be consistent with a primary of somewhat less than 15 \msun\ accreting between 1 and 10 \msun\ in the CHeB and CCB epochs. The observed equatorial velocity might also be attained by accreting a broad range of masses onto a primary of somewhat more than 20 \msun\ in the later CSB epoch. }  Although our treatment of the post-merger problem with \textsc{mesa} is rather different than that of \citet{chatz20}, we note that the results for the final equatorial rotational velocity (their Table 1) are very similar to ours. This gives us some confidence that this quantity is somewhat robust against the details of the merger process and depends primarily on a global quantity such as pre-merger orbital angular momentum.

For our study to have any relevance to Betelgeuse, it is important that the structure remain that of an RSG after the proposed merger. As mentioned in \S \ref{comp}, a ``quiet merger" can leave behind an RSG, depending on pre-merger conditions. \citet{IvanovaPod03} suggest that this condition favors secondary masses $>2$ \msun\ and a primary close to carbon ignition so that strong gradients inhibit core/envelope mixing. 

To account for the circumstellar nebular rings, many studies of the mergers of massive stars have focused on the prospect that the progenitor of SN~1987A may have undergone a merger \citep{MorrisPod07}. Merger models can also account for why the progenitor was a blue rather than red supergiant by invoking mixing of helium from the core into the outer envelope \citep{MenonHeger17}. Mixing can happen when the secondary nears the helium core, fills its Roche lobe, and produces a mass transfer stream that can penetrate the core \citep{IPS02}. The depth of penetration of the stream into the core depends on the stream direction, entropy, width, and angular momentum, the rotation orientation and amplitude of the secondary, on the density structure and relative rotation of the core, and on fluid instabilities. In the case of Betelgeuse, a contrasting conclusion applies. Betelgeuse is still a red supergiant. If one accepts our basic {\it ansatz} that a merger is required to account for the observed rotational velocity of Betelgeuse, then it follows that a merger did not produce a compact blue envelope and thus, by the arguments of \citet{IPS02} and \citet{MenonHeger17}, little to no helium could have been mixed outward from the core, consistent with our particular simulations. The modeling of a putative Betelgeuse merger by \citet{chatz20}  concluded that the plume from the disrupted secondary would not penetrate the helium core and induce substantial helium mixing according to the prescription of \citet{IPS02}. Mixing may be more likely for more massive secondaries, so our results may be less reliable for larger mass secondaries. Plume mixing is a complex hydrodynamical problem that deserves more study if we are to understand both Betelgeuse and SN~1987A as products of massive star mergers.

For our accreting models, a wave of angular momentum is halted at the composition boundary at the edge of the helium core leaving behind an envelope of constant angular velocity and a monotonically rising angular momentum per unit mass. \green{The composition distribution of the inner core can be slightly affected by the accretion of a companion of large mass. Accretion has little effect on the production of magnetic fields by the Spruit/Tayler mechanism.} 
 
Thus while the inner structure might be somewhat perturbed by accretion of substantial mass, there may be little on the outside to indicate that the accretion occurred. \green{Our models provide a reasonable ``natural" explanation for why Betelgeuse has a large, but sub-Keplerian equatorial velocity.} Our results do not prove, but do allow that Betelgeuse might have merged with a moderately massive companion. Betelgeuse might look substantially the same whether it merged with a 1 or 10 \msun\ companion.
 
While we have run all of our models to near core collapse and examined the conditions there, the pertinent question is the structure and condition of Betelgeuse as we see it today, gracing Orion. While uncertainties in the distance remain troubling, Betelgeuse is most likely near the tip of the RSB. Since core helium burning is far longer than subsequent burning phases, Betelgeuse is most likely in core helium burning, a point reinforced by \citet{Joyce20}. For our models, once the transient phase of accretion has settled down and substantial mass and angular momentum have been ejected, there is rather little external difference in models in late core helium burning and subsequent phases.

In \citet{Wheeler17}, we noted that a merger event might have some relation with the interstellar shells of higher density in the vicinity of Betelgeuse. The strangely linear feature about 9' away might be related to the square axisymmetric circumstellar nebula recently discovered around the B9 Ia star HD93795 by \citet{Gvar20}. The prominent bow shock at $\sim 7'$ in the same direction indicates a peculiar velocity with respect to the local standard of rest of $v \approx 25$ \kms\ \citep{2008AJ....135.1430H} or perhaps as much as 35 \kms\ \citep{vanloon2013}. This number is of relevance because some hypothesize that this rather high peculiar velocity arises because Betelgeuse is a runaway star, having been ejected when a binary companion underwent a supernova explosion \citep{blaauw61,vanloon2013}. If a previous binary companion exploded, then it clearly could not have merged with the current Betelgeuse. Work on the kinematic effects of supernovae in massive star binary systems tends to discourage this conjecture. \citet{Renzo19} confirm that of order 20 - 50\% of massive star binaries merge, as we explore here. They also find that by far the largest fraction of binaries disrupted by the collapse and explosion of the primary result in ``walkaway" rather than ``runaway" stars. The velocity distribution of the ejected companion peaks at about 6 \kms. For secondaries more massive than 15 \msun, as likely applies to Betelgeuse, only $\sim 0.5\%$ have velocities of 30 \kms\ and above, as appropriate to Betelgeuse. These results suggest that, while non-zero,
the likelihood that the space motion of Betelgeuse resulted from the previous explosion of a companion is small. An alternative is that the proper motion of Betelgeuse arises from stellar dynamics in its natal cluster \citep{poveda67,ohkroupa16,schoettler19}. The results depend on assumptions about primordial binaries, among other things, but the general results are roughly the same. It is easier to generate walkaway stars than runaway stars. A runaway binary is likely to be rare, but not precluded. 

The origin of the space motion of Betelgeuse is thus one more fascinating open question about this tantalizing star. Whether Betelgeuse attained its proper motion from the explosion of a companion or from cluster dynamics, if it emerged as a single star then the observed equatorial velocity remains an issue. Even if spun up on the ZAMs, its rotation on the RSB would be slow \citep{Wheeler17}. A possible way to account for both the equatorial velocity and the space motion would be to invoke cluster dynamics, ejection of a binary of which the star we currently observe as Betelgeuse was the primary, and a subsequent merger to account for the equatorial velocity. This is, admittedly, an improbable string of events. \citet{ohkroupa16} find that a majority of ejected massive binaries have a period shorter than
$10^5$ days. Our merger models have a typical presumed orbital period of about 30 years or $10^4$ days. Having a rather massive companion might increase the likelihood that the binary remains intact upon ejection from the natal cluster. Our current results allow for that possibility. We note that while Betelgeuse may have moved hundreds of pc during its main sequence lifetime, it is expected to have moved only $\sim 2$ pc during the 100,000 years or so it has been in core helium burning as a RSG.

While the overall goal of the Betelgeuse Project is to determine the current evolutionary phase and ultimate fate of Betelgeuse, this work has brought us no closer to a practical observational test of those important aspects. The notion that Betelgeuse may have undergone a merger remains viable.

\section*{Acknowledgments}

We are grateful to Natasha Ivanova for discussions of common envelope evolution and to the Aspen Center for Physics for providing the environment to do so. We also thank Manos Chatzopoulos, Juhan Frank, Meridith Joyce, and Andrea Dupree and the Month of Betelgeuse (MOB) team for discussions of Betelgeuse and mergers. We are especially thankful for the ample support of Bill Paxton and the MESA team. This research was supported in part by the Samuel T. and Fern Yanagisawa Regents Professorship in Astronomy and by NSF AST-1109801 and NSF AST-1813825.

\software{MESA (Paxton et al. 2011, 2013, 2015, 2018)}. 


\end{document}